\documentclass{elsart}
\usepackage{graphicx,natbib,amssymb}
\journal{New Astronomy}

\usepackage{graphics}
\usepackage{rotating}
\usepackage{multirow}

\newcommand{\apj}{ApJ\phantom{s}}

\newcommand{\apjs}{ApJSS\phantom{s}}

\newcommand{\araa}{ARA\&A\phantom{s}}

\def\Mesz{M\'esz\'aros}

\def\beq{ \begin{equation} }
  \def\eeq{ \end{equation} }
\def\beqa{ \begin{eqnarray} }
  \def\eeqa{ \end{eqnarray} }

\def\astrobj#1{#1}

\begin{document}
\begin{frontmatter}

  \title{Diffuse high energy neutrinos
    and cosmic rays from hyperflares of soft-gamma repeaters}

  \author[label1,label2,label3]{Xue-Wen Liu}
  \ead{xwliu@pmo.ac.cn},
  \author[label1,label3,label4]{Xue-Feng Wu}
  \ead{xfwu@pmo.ac.cn},
  \author[label1,label3]{Tan Lu}
  \ead{t.lu@pmo.ac.cn}

  \address[label1]{Purple Mountain Observatory, Chinese
    Academy of Sciences, Nanjing, 210008, China}
  \address[label2]{Graduate School, Chinese Academy of Sciences,
  Beijing, 100039, China}
  \address[label3]{Joint Center for Particle Nuclear Physics
    and Cosmology (J-CPNPC), Nanjing 210093, China}
  \address[label4]{Department of Astronomy $\&$ Astrophysics,
    525 Davey Lab, Pennsylvania State University, University
    Park, PA 16802, USA}

  \label{firstpage}

  \begin{abstract}
    We calculate the diffuse high energy (TeV - PeV) neutrino emission
    from hyperflares of Soft-Gamma Repeaters (SGRs), like the hyperflare risen from \astrobj{SGR
    1806-20} on December 27 of 2004, within the framework of the fireball
    model. The fireball model for gamma-ray bursts (GRBs) can explain
    well the main features of this hyperflare and the subsequent
    multi-frequency afterglow emission.
    The expected rate, $\sim 20-100$
    Gpc$^{-3}$day$^{-1}$, of such hyperflares is well in excess of the GRBs rate. Our result
    shows that the contribution to
    the diffuse TeV-PeV neutrino background from such hyperflares
    is less than $10\%$ of the contribution from GRBs. We
    also discuss the high energy cosmic rays (CRs) from these
    sources.
  \end{abstract}

\end{frontmatter}

\section{Introduction}\label{sec:intro}
It is a general consensus that relativistic shocks can accelerate
nuclei to very high energies through the Fermi acceleration
mechanism. By interacting with photons or baryons, such high
energy nuclei (mostly protons) can generate pions and the latter
cascade into neutrinos and leptons. So, for a relativistic jet
that sweeps up its surrounding medium, it could be an efficient
high-energy neutrino producer, if the jet is also surrounded by
intense photon fields. Three famous kinds of such astronomical
objects are gamma-ray bursts (GRBs; Waxman 1995; Vietri 1995;
Waxman \& Bahcall 1997; Bahcall \& \Mesz 2000; \Mesz \& Waxman
2001; Guetta, Spada \& Waxman 2001; Dermer \& Atoyan 2003; Asano
\& \Mesz 2008), active galactic nuclei (Axford 1994; Atoyan \&
Dermer 2001; Dermer, Ramirez-Ruiz \& Le 2007; Berezhko 2008;
Abbasi et al. 2008), and micro-quasars (Levinson \& Waxman
2001; Distefano et al. 2002), all of which are usually discussed as
high-energy neutrino and cosmic ray (CR) sources. For some
thorough reviews on astrophysical neutrinos and their connection
to CRs, we would like to refer the readers to the references
Halzen et al. (2002) and Becker (2008). In this paper, we would
focus on another kind of such astronomical objects, i.e.,
soft-gamma repeaters (SGRs) which are widely accepted as magnetars
(pulsars with super strong magnetic field $\sim 10^{15}\rm{G}$;
Duncan \& Thompson 1992). Zhang et al. (2003) proposed a model for
neutrino production by magnetars in their steady phase of periodic
emission. In contrast, some short neutrino
bursts could also be produced by a relativistic outflow during the
violent giant-flare phase of SGRs.

Giant flares are distinguished from common SGR bursts by their
extreme energies ($\rm{\sim10^{44}~ergs}$) emitted during their
initial short ($\sim 0.1~\rm{s} -0.5~\rm{s}$) pulses of soft gamma
rays followed by subsequent emission lasting hundreds of seconds
showing pulsations associated with the spinning neutron star
(Woods \& Thompson 2006). Among the observed three giant flares
during the last four decades, the brightest one originated from
the \astrobj{SGR1806-20} on 2004 December 27 (Hurley et al. 2005;
Palmer et al. 2005; Terasawa et al. 2005; Mazets et al. 2005) has
an energy release exceeding $10^{46}$ ergs, which is two orders of
magnitude higher than the energy release of the other two. The
huge difference in the luminosity urges us to consider this kind
of events separately and name it "hyperflare" (Popov \&
Stern 2006). The mechanism triggering hyperflares remains a matter of
debate. Lugones (2007) proposed a model in which the core
conversion of an isolated neutron star with a magnetic field of
$\sim 10^{12}$ G and a fallback disk around it can trigger a giant
flare. In the popular magnetar model, giant flares result from a
global magnetic rearrangement within the crust of the magnetar
(Thompson \& Duncan 2001). Since giant flares have many
similarities to GRBs which can be well understood by the fireball
model (Piran 1998; \Mesz 2002), different scenarios have been
proposed within the frame work of the fireball model (Wang et al.
2005; Dai et al. 2005; Yamazaki et al. 2005), to explain the
abundant multi-frequency afterglow data of this hyperflare
(Gaensler et al. 2005; Cameron et al. 2005; Taylor et al. 2005)
and the flare itself might be the emission from the internal shock
and/or the photosphere of the fireball if the relativistic outflow
is variable (Nakar, Piran \& Sari 2005; Ioka et al. 2005). Ioka et
al. (2005) estimated the number of high-energy neutrino events
from this hyperflare and argued that the neutrino flux should be
detected by current neutrino observatories such as AMANDA (Ahrens
et al. 2002; Halzen, Landsman \& Montaruli 2005), which may put
constraints on the flare mechanism. Fan, Zhang \& Wei (2005)
considered the production of neutrinos with typical energy of
$10^{14}$ eV through photomeson interaction of X-ray tail photons
with $\sim10^{17}$ eV CRs accelerated in the external forward
shock by this hyperflare. They found that the neutrino fluence
produced in the external shock is too weak to be detected.

The rate of hyperflares is very uncertain because of low
statistics based on a small sample. Their intrinsic low
release energies relative to GRBs make them impossible to be
detected beyond $\sim 30-40$ Mpc by BATSE and $\sim$ 70 Mpc
by Swift (Hurley et al.
2005). Lazzati et al. (2005) gave a slightly less stringent limit
for the rate of hyperflares: $< 1/130~ \rm yr^{-1}$ per a
Milky-Way-like galaxy. Popov \& Stern (2006) argued that the rate
is $\sim 10^{-3}~\rm yr^{-1}$. They further conservatively
estimated the expected rate of hyperflares to be $\sim 20-100 ~\rm
Gpc^{-3}~day^{-1}$ (Popov \& Postnov 2007). Recently, Lorimer et
al. (2007) reported a discovery of a strong millisecond
extragalactic radio burst with peculiar properties and estimated
that the cosmological rate for this radio burst is $\sim 50~\rm
Gpc^{-3}~day^{-1}$ which is in correspondence with the
statistically estimated rate of hyperflares. Meanwhile, the
millisecond time scale of the radio burst is consistent with an
event in the magnetosphere of a magnetar, indicating that both the
millisecond extragalactic radio burst and hyperflares may possibly
come from the same source: magnetars (Popov \& Postnov 2007).

The above rate of hyperflares in the universe is well in excess of
the GRBs rate, while the total energy release of a hyperflare is
much lower than that of a typical GRB. Based on the estimated rate
and the fireball model proposed by Ioka et al. (2005) for the
hyperflare of \astrobj{SGR 1806-20} (see Section 2) in which
relativistic protons are accelerated by an internal shock and
target photons are dominated by the hyperflare, we describe the
neutrino production process within hyperflares in Section 3. We
estimate the diffuse TeV-PeV neutrino flux from hyperflares,
compare our results with that from magnetar steady phase and from
GRBs in Section 4. We also give a discussion on high-energy CRs
from hyperflares in this section. We summarize our results and
conclusions in Section 5.

\section{The fireball model for the hyperflare from \astrobj{SGR 1806-20}}
The hyperflare of \astrobj{SGR 1806-20} is the only event up to
now and its spectrum may be either thermal (Hurley et al. 2005) or
non-thermal (Mazets et al. 2005; Palmer et al. 2005). The baryon
load in the outflow is less constrained in contrast to GRBs, then
the outflow produced the hyperflare may be either baryon-poor or
baryon-rich (Ioka et al. 2005). Based on the fireball model, we
describe these two scenarios as follows.

The isotropic soft $\gamma-$ray energy of the December 27 event,
$E_{\gamma}\sim 3\times 10^{46} E_{\gamma,46.5}$ ergs, released
within a time interval of $t_0\sim 0.1t_{0,-1}$ s from somewhere
near the surface of the magnetar with radius of $r_0\sim 10^6$ cm,
could create an optically thick fireball with an initial
temperature of
\begin{equation}
  T_0 \sim
  \Big(\frac{L_0}{4\pi r_0^2ca}\Big)^{1/4}\sim
  300~L_{0,47.5}^{1/4}~r_{0,6}^{-1/2}~{\rm keV},
\end{equation}
where $a$ is the radiation density constant, $L_0\sim
L_{\gamma}/\xi_{\gamma}\sim 3\times 10^{47}L_{0,47.5} $ ergs ${\rm
s^{-1}}$ and $\xi_{\gamma}$ is the conversion efficiency of total
energy into gamma-rays (Ioka et al. 2005). The subsequent fireball
evolution depends on the dimensionless entropy $\eta=L_0/\dot M
c^2$ which has a critical value (\Mesz \& Rees 2000)
\begin{equation}
  \eta_{\ast}=\Big(\frac{L_0\sigma_T}{4\pi
    m_pc^3r_0}\Big)^{1/4}\sim
  100~L_{0,47.5}^{1/4}~r_{0,6}^{-1/4}.
\end{equation}
If $\eta<\eta_{\ast}$, the fireball is baryon-rich, then the
photosphere appears in the coasting phase and almost all the
energy goes into the kinetic luminosity of the outflow, $L_{{\rm
kin}}\sim L_0$. While if $\eta>\eta_{\ast}$, the fireball is
baryon-poor, the photosphere appears in the acceleration phase and
then a small fraction of the energy goes into the kinetic
luminosity of the outflow $L_{\rm kin}\sim L_\gamma\times
\eta_{\ast}/\eta$ (Ioka et al. 2005). The above two scenarios are
both possible for the December 27 hyperflare. The particular
parameters adopted for the two scenarios by Ioka et al. (2005) are
listed below

\textbf{Baryon Poor}: $\eta\sim 10^4$, $L_{\rm kin}\sim
10^{-2}L_{\gamma}\sim 10^{45.5}$ ergs ${\rm s^{-1}}$,
$\Gamma\sim 100$, $\Delta t\sim 10^{-4}$ s,

\textbf{Baryon Rich}: $\eta\sim 10$, $L_{\rm kin}\sim
10L_{\gamma}\sim 10^{48.5}$ ergs ${\rm s^{-1}}$, $\Gamma\sim
10$, $\Delta t\sim 10^{-1}$ s.

where $\Gamma$ is the Lorentz factor of the outflow and $\Delta t$
is the variability timescale of the hyperflare. It should be noted
that the photosphere emission is thermal. If the internal shocks
occur above the photosphere as in the baryon-rich model, the
non-thermal shock emission will dominate the photosphere thermal
emission if $\eta<100\eta_{\ast,2}\xi_s^{3/8}$, where $\xi_s$ is
the conversion efficiency of kinetic energy into photons (Ioka et
al. 2005).

\section{The neutrino production in hyperflares}
The photon distribution in the comoving frame of the
internal shocked region is usually assumed to be isotropic
and described by monoenergetic number density $dn_\gamma/d\epsilon_\gamma$ at
energy $\epsilon_\gamma$. The protons are usually
assumed to be accelerated to a power law distribution by
internal shocks, $dn_p/d\epsilon_p\propto \epsilon_p^{-2}$ where
the proportional coefficient $1/{{\rm
  ln}(\epsilon_{p,\rm max}/\epsilon_{p,\rm min})}$ as the fraction of the total
energy that is contributed by each decade of energy is about 0.1.
When bathed in the hyperflare photon field, the high-energy
protons would lose their energy by $p\gamma$ interaction,
resulting in plenty of pions. Due to the pion production, the
fractional energy loss rate of a proton with energy
$\epsilon_p=\gamma_pm_pc^2$ is (Stecker 1968; Waxman \& Bahcall 1997)
\begin{equation}
  t_{p\gamma}^{-1}=-\frac{1}{\epsilon_p}\frac{d\epsilon_p}{dt}=\frac{c}{2\gamma_p^2}
  \int_{\bar \epsilon_{\gamma,\rm
      th}}^{\infty}d\bar{\epsilon}_\gamma
  \bar{\sigma}_{p\gamma}(\bar{\epsilon}_\gamma)\bar{\kappa}_{p\gamma}
  (\bar{\epsilon}_\gamma)\bar{\epsilon}_\gamma\int_{\bar{\epsilon}_{\gamma}/2
    \gamma_p}^{\infty}d\epsilon_\gamma
  \epsilon_\gamma^{-2}\frac{dn_\gamma}{d\epsilon_\gamma}
\end{equation}
where $\bar{\sigma}_{p\gamma}(\bar {\epsilon}_\gamma)$ is the cross section
of the photopion interaction for a target photon with energy
$\bar{\epsilon}_\gamma$ in the proton rest frame, $\bar{\kappa}_{p\gamma}$ is
the inelasticity coefficient defined as the average fraction of
energy lost to the pion and $\bar \epsilon_{\gamma, \rm th}=0.15$ GeV is
the threshold energy. Using the $\Delta$-resonance
approximation, Eq. (3) can be casted into
\begin{equation}
  t_{p\gamma}^{-1}\approx \frac{c\bar{\sigma}_{\rm
      peak}\bar{\kappa}_{\rm peak}
    \bar{\epsilon}_{\Delta}
    \overline{\Delta\epsilon}}{2\gamma_p^2} \int_{\bar{\epsilon}_{\Delta}/2
    \gamma_p}^{\infty}d\epsilon_\gamma \epsilon_\gamma^{-2} \frac{dn_\gamma}{d\epsilon_\gamma}
\end{equation}
where $\bar{\sigma}_{\rm peak}\simeq 5\times 10^{-28}$ ${\rm
  cm^2}$ and $\bar{\kappa}_{\rm peak}\simeq 0.2$ are the values of
$\bar{\sigma}$ and $\bar{\kappa}$ at
$\bar{\epsilon}_\gamma=\bar{\epsilon}_{\Delta}=0.3$ GeV where the
cross section peaks due to the $\Delta$ resonance, and
$\overline{\Delta\epsilon}\simeq 0.2$ GeV is the peak width.

Under the $\Delta$-resonance approximation, the neutrino
spectrum is totally determined by the parameter $f_{p\gamma}\simeq r_{\rm
  sh}/\Gamma c t_{p\gamma}$, the fraction of
energy lost by protons to pions, where the internal shock radius
$r_{\rm sh}$ is $\sim 2\Gamma^2c\Delta t$. At each $p\gamma$
interaction, a proton loses $\sim 20\%$ of its energy which is
distributed roughly equally among the products of the decay
processes $\pi^{\pm}\rightarrow\mu^{\pm}+\nu_{\mu}(\bar
\nu_{\mu})\rightarrow
e^{\pm}+\nu_e(\bar\nu_e)+\nu_{\mu}+\bar\nu_{\mu}$. In addition the
neutrino oscillation will change neutrino flavor from
$\nu_e:\nu_\mu:\nu_\tau=1:2:0$ at the source site to $1:1:1$ at
the earth, which will reduce the observed muon neutrino flux by a
factor of 2. Once we know the spectrum of the target photon, we
can calculate the monoenergetic muon neutrino flux as
\begin{equation}
  \Phi_{\nu,p\gamma}={\rm min}(1,f_{p\gamma})\frac{1}{8}\frac{\xi_i
    L_{\rm kin}}{4\pi d^2\epsilon_{\nu,\rm typ}^2},
\end{equation}
where $d$ is the luminosity distance of the source, $\xi_i$ is the
fraction of the total kinetic energy converted into accelerated
protons and $\epsilon_{\nu,\rm
  typ}$ is the typical energy of neutrinos produced by the
$p\gamma$ interactions. Below we consider two different types of
photon spectra of SGR hyperflares.

\subsection{The thermal spectrum of hyperflares}
Hurley et al. (2005) reported that the observed energy spectrum of
the hard spike of the December 27 hyperflare is consistent with a
cooling blackbody with average temperature $k_B
T_{\rm{obs}}=175\pm25$ keV, and thus the photon peak energy in the
observer frame is $\epsilon_{\gamma,\rm peak}^{\rm{obs}}\approx
2.7k_B T_{\rm{obs}}$. The differential photon field density in the
comoving frame of the internal shocked region is therefore
$dn_\gamma/d\epsilon_\gamma=({8\pi}/{h^3c^3})({\epsilon_\gamma^2}/
({e^{\epsilon_\gamma/k_BT}-1}))$, where $T=T_{\rm{obs}}/\Gamma$
and $\epsilon_\gamma=\epsilon_\gamma^{\rm obs}/\Gamma$. From
$n_{\gamma}=\int _0
^{\infty}\frac{dn_\gamma}{d\epsilon_\gamma}d\epsilon_\gamma\simeq
L_\gamma/(4\pi r^2 c \Gamma \epsilon_{\gamma,\rm
peak}^{\rm{obs}})$, we have (Ioka et al. 2005)
\begin{eqnarray}
  n_\gamma\sim \cases{7\times
    10^{18}~L_{\gamma,47.5}~\epsilon_{\gamma,5.3}^{-1}~\Gamma_2^{-5}~\Delta t_{-4}^{-2} ~~{\rm cm^{-3}}
    ~~({\rm Baryon~ Poor})\cr 7\times
    10^{17}~L_{\gamma,47.5}~\epsilon_{\gamma,5.3}^{-1}~\Gamma_1^{-5}~\Delta t_{-1}^{-2} ~~{\rm cm^{-3}}
    ~~({\rm Baryon~ Rich})},
\end{eqnarray}
where $\epsilon_{\gamma,5.3}=\epsilon_{\gamma,\rm peak}^{\rm{obs}}/200$
keV.

Inserting the differential photon number density into Eq. (4), we
obtain the analytical form of the fractional energy loss rate as
\begin{equation}
  t_{p\gamma}^{-1}\approx
  \frac{U_{\gamma}c}{\epsilon_{\gamma,\rm peak}}\bar{\sigma}_{\rm
    peak}\bar{\kappa}_{\rm
    peak}\frac{\overline{\Delta{\epsilon}}}{{\bar{\epsilon}}_{\Delta}}\times x^2\Big|{\rm
    ln}(1-e^{-x})\Big|,
\end{equation}
where the dimensionless parameter
$x={\bar{\epsilon}}_{\Delta}/(2\gamma_pk_BT)$ and the photon
energy density
$U_{\gamma}=n_{\gamma}\epsilon_{\gamma,\rm{peak}}=L_{\gamma}/(4\pi
r^2 c \Gamma^2)$. Then $f_{p\gamma}$ can be derived analytically,
\begin{eqnarray}
  f_{p\gamma,\rm th}&\simeq& x^2\Big|{\rm ln}(1-e^{-x})\Big| \nonumber\\
  && \times
  \cases{0.28~L_{\gamma,47.5}~\epsilon_{\gamma,5.3}^{-1}~\Gamma_2^{-4}~\Delta t_{-4}^{-1}
    ~~({\rm Baryon~ Poor})\cr
    2.8~L_{\gamma,47.5}~\epsilon_{\gamma,5.3}^{-1}~\Gamma_1^{-4}~\Delta t_{-1}^{-1}
    ~~~~({\rm Baryon~ Rich})},
\end{eqnarray}
which peaks at $x_{\rm
  peak}=2.7\bar{\kappa}\Gamma^2\bar{\epsilon}_\Delta
m_pc^2/(8\epsilon_{\nu}^{\rm{obs}} \epsilon_{\gamma,\rm
  peak}^{\rm{obs}})\approx 1.8$. Approximation of the scaling
function $|{\rm ln}(1-e^{-x})|$ yields $e^{-x}$ for $x>x_{\rm
  peak}$, and ${\rm |ln}x|$ for $x\ll 1$. Thus, the exponential
suppression of the neutrino spectrum appears for
$\epsilon_{\nu}^{\rm{obs}}<{3\bar{\kappa}\Gamma^2\bar{\epsilon}_\Delta
  m_pc^2}/(16\epsilon_{\gamma,\rm{peak}}^{\rm{obs}})$, and high energy
  region of neutrinos scales as $x^2|{\rm ln}x|$. Meanwhile, we get the peak energy of the
neutrino spectrum
\begin{eqnarray}
  \epsilon_{\nu,\rm{peak}}^{\rm{obs}}=\frac{3\bar{\kappa}\Gamma^2\bar{\epsilon}_\Delta
    m_pc^2}{16\epsilon_{\gamma,\rm{peak}}^{\rm{obs}}}\sim\cases{
    560~\Gamma_2^2~\epsilon_{\gamma,5.3}^{-1}~~ \rm TeV, ~~(Baryon~ Poor) \cr
    5.6~\Gamma_1^2~\epsilon_{\gamma,5.3}^{-1}~~ \rm TeV, \;
    ~~(Baryon~ Rich)}.
\end{eqnarray}

\subsection{The non-thermal spectrum of hyperflares}
The photon spectrum of the 2004 December 27 hyperflare is also
likely to be non-thermal, making this event resemble a short, hard
gamma-ray burst (SHB). Palmer et al. (2005) argued that such an
energetic SGR flare may indeed form a subclass of GRBs. With a
luminosity of $\sim 10^{47}$ ergs ${\rm
  s^{-1}}$, such flares can be detected by BATSE up to $\approx$ 40
Mpc, suggesting that a considerable fraction of the BATSE SHB
sample is compromised of similar extragalactic hyperflares (Palmer
et al. 2005; Hurley et al. 2005; Popov \& Stern 2006; Nakar 2007).
Another two short hard GRBs, 051103 and 070201, are recently
reported to show evidences as hyperflares from SGRs in the nearby
M81 and M31 galaxies, respectively (Frederiks et al. 2007; Mazets
et al. 2008). So although very uncertain, at least part of
hyperflares from SGRs would have a similar spectrum to SHBs. Since
the non-thermal component is negligible in the baryon-poor model,
we only consider a typical non-thermal hyperflare spectrum in the
baryon-rich scenario with a break energy
$\epsilon_{\gamma,b}^{\rm{obs}}\sim 200$ keV
\begin{eqnarray}
  \frac{dn_\gamma}{d\epsilon_\gamma}=n_b\times \cases{
    \Big(\frac{\epsilon_\gamma}{\epsilon_{\gamma,b}}\Big)^{-\alpha},
    ~~\epsilon_\gamma<\epsilon_{\gamma,b}\cr
    \Big(\frac{\epsilon_\gamma}{\epsilon_{\gamma,b}}\Big)^{-\beta}, ~~\epsilon_\gamma>
    \epsilon_{\gamma,b}},
\end{eqnarray}
where $\epsilon_{\gamma,b}=\epsilon_{\gamma,b}^{\rm
  obs}/\Gamma$ and $n_b\sim U_{\gamma}/2{\epsilon_{\gamma,b}}^2$. After performing the
integration in Eq. (4), we can approximate the energy loss rate of
protons by (Waxman \& Bahcall 1997; Murase \& Nagataki 2006)
\begin{eqnarray}
  t_{p\gamma}^{-1}=\frac{U_{\gamma}c}{2\epsilon_{\gamma,b}}\bar{\sigma}_{\rm
    peak}\bar{\kappa}_{\rm peak}
  \frac{\overline{\Delta{\epsilon}}}{{\bar{\epsilon}}_{\Delta}}\times
  \cases{
    \Big(\frac{{\bar{\epsilon}}_{\Delta}}{2\gamma_p\epsilon_{\gamma,b}}\Big)^{1-\alpha},
    ~~\epsilon_\gamma<{\epsilon_{\gamma,b}}\cr
    \Big(\frac{{\bar{\epsilon}}_{\Delta}}{2\gamma_p\epsilon_{\gamma,b}}\Big)^{1-\beta},
    ~~\epsilon_\gamma>{\epsilon_{\gamma,b}}},
\end{eqnarray}
and thus
\begin{eqnarray}
  f_{p\gamma,\rm non-th}&\simeq&
  1.4~L_{\gamma,47.5}~\epsilon_{\gamma,5.3}^{-1}~\Gamma_1^{-4}~\Delta
  t_{-1}^{-1}
  \cases{(\frac{\epsilon_{\nu}^{\rm{obs}}}{\epsilon_{\nu,b}^{\rm{obs}}}\Big)^{\beta-1},
    ~~\epsilon_{\nu}^{\rm{obs}}<\epsilon_{\nu,b}^{\rm{obs}}\cr
    (\frac{\epsilon_{\nu}^{\rm{obs}}}{\epsilon_{\nu,b}^{\rm{obs}}}\Big)^{\alpha-1},
    ~~\epsilon_{\nu}^{\rm{obs}}>\epsilon_{\nu,b}^{\rm{obs}}}
\end{eqnarray}
where $\epsilon_{\nu,b}^{\rm{obs}}\approx
\bar{\kappa}\Gamma^2\bar{\epsilon}_\Delta
m_pc^2/(8\epsilon_{\gamma,b}^{\rm{obs}})\approx 4\Gamma_{1}^2$ TeV
is the neutrino break energy in the observer frame. The power law
neutrino spectrum resulting from a non-thermal target photon field
is quite different to the one predicted from a thermal photon
field as in Eq. (8).

\section{The diffuse high-energy neutrinos and CRs from hyperflares}
Similar to GRBs, a high-energy neutrino flash accompanying an SGR
hyperflare would be expected if the source is nearby and
energetic. Below we discuss the probability of detecting TeV-EeV
muon neutrinos from the \astrobj{SGR 1806-20} hyperflare by
IceCube, using the following formula (Razzaque et al. 2004; Ioka
et al. 2005)
\begin{equation}
  P(\epsilon_\nu)=7\times 10^{-5}(\epsilon_\nu/10^{4.5}~\rm GeV)^{\beta},
\end{equation}
where $\beta=1.35$ for $\epsilon_\nu<10^{4.5}~\rm GeV$ while
$\beta=0.55$ for $\epsilon_\nu>10^{4.5}~\rm GeV$. The number of
muon events from muon neutrinos above TeV is given by
\begin{equation}
  N_\mu=A_{\rm det}t_0\int_{\rm TeV}^{\rm PeV}P(\epsilon_\nu)\Phi_{\nu,p\gamma}d\epsilon_\nu,
\end{equation}
where the geometrical detector area of IceCube $A_{\rm
  det}\sim 1 ~\rm km^2$ ($A_{\rm det}\sim 0.03 ~\rm
km^2$ for AMANDA). Then the number of muon events from
$p\gamma$ neutrinos at typical energy ($\epsilon_{\nu,\rm
  typ}=\epsilon_{\nu,\rm peak}^{\rm obs}$ for thermal spectrum and $\epsilon_{\nu,\rm
  typ}=\epsilon_{\nu,b}^{\rm obs}$ for non-thermal spectrum) is
\begin{eqnarray}\label{eq:numberofmuon}
  N_{\mu}&=&A_{\rm det}t_0P(\epsilon_{\nu,\rm
    typ})\epsilon_{\nu,\rm
    typ}\Phi_{\nu,p\gamma}\nonumber\\
  &\sim&\cases{1.3~{\rm min}(1,f_{p\gamma,\rm th})\xi_{i,-1}L_{\rm
      kin,45.5}d_{1}^{-2}(\epsilon_{\gamma,5.3}\Gamma_2^2)^{\beta-1}
    A_{\rm det,1}t_{0,-1}~(\rm Baryon~ Poor)\cr
    2500~{\rm min}(1,f_{p\gamma,\rm th})\xi_{i,-1}L_{\rm kin, 48.5}
    d_{1}^{-2}(\epsilon_{\gamma,5.3}\Gamma_1^2)^{\beta-1}A_{\rm det,1}t_{0,-1}~(\rm Baryon ~Rich)\cr
    2200~{\rm min}(1,f_{p\gamma,\rm non-th})\xi_{i,-1}L_{\rm kin, 48.5}
    d_{1}^{-2}(\Gamma_1^2)^{\beta-1}A_{\rm det,1}t_{0,-1}~~(\rm Baryon ~Rich)
  },
\end{eqnarray}
where the distance of \astrobj{SGR 1806-20} is taken
to be $10d_{1}~\rm kpc$ (see Ioka et al. 2005; Dai et
al. 2005; however $d\sim15$ kpc in Corbel \& Eikenberry 2004).
Due to the model uncertainties, the number of muon events spans
more than three orders of magnitude. To see clearly the behavior
of the number of muon events depending on the model parameters, we
can fix the values of $\xi_i$, $d$ and set $\Delta t$ free in the
baryon-poor model, then the event number is approximately
proportional to $\eta_\ast/\eta({\rm
  min}[\eta,\eta_\ast])^{2\beta-2}$. Parameter space in the $\eta - \Delta t$
plot where IceCube can detect more than one muon event from SGR
1860-20 is shown in Fig. 1 of Ioka et al. (2005). It can be seen that a
baryon-poor outflow with entropy less than 2500 can trigger one
event. While if the outflow is baryon-rich, IceCube can detect
about one event even when a hyperflare considered here is located
$\sim 200~\rm kpc$ away. If such detection comes true, it will
provide independent evidence for the picture in which relativistic
outflows produce hyperflare electromagnetic emission and constrain
the parameters of the baryon loading, the bulk Lorentz factor of
the fireball, the efficiency of energy conversion and the
variability timescale of the hyperflare. It should be noted that
we neglect the neutrino production through the $pp$ reaction in
this paper.

Now we turn to the observations. The AMANDA-II detector
was running to search for down-going muons cascading into
high-energy gamma-rays and neutrinos, when the hyperflare on
December 27, 2004 saturated many satellite gamma-ray detectors.
However, the data revealed no significant signal which put an upper
limit on neutrino flux of the hyperflare: $\Phi_\nu < 0.4(6.1)
~\rm TeV ~m^{-2}~s^{-1}$ for an energy spectrum index $-1.47~(-2)$
(Achterberg et al. 2006). This limit would suggest a baryon-poor
outflow, which means that the diffuse neutrino flux from the
hyperflare is much lower than $1.0\times 10^{-10}~\rm
GeV~cm^{-2}~s^{-1}$, far below the IceCube sensitivity 
$\sim 8.0\times 10^{-9}~\rm GeV~cm^{-2}~s^{-1}$  after one year (Ahrens
et al. 2004; Hoshina et al. 2008; Halzen 2008). But IceCube,
currently under construction at the South pole, can potentially
detect TeV-PeV neutrinos from a single hyperflare originating from
a baryon-poor outflow (see Eq. \ref{eq:numberofmuon}). Another
detector Km3NeT, a planned experiment in the Mediterranian Sea to
complement the IceCube, will have a better detectability for a
south-hemisphere source like \astrobj{SGR 1806-20} due to its low
background of atmospheric muons (Katz 2006). As estimated above,
in the baryon-rich model, a hyperflare with the same energy as the
hyperflare of \astrobj{SGR 1806-20} can be detected by IceCube
within $\sim 200~\rm kpc$. If AMANDA-II really had not observed
high energy neutrinos from \astrobj{SGR 1806-20}, then IceCube can
only detect hyperflares in our galaxy. \astrobj{SGR 1806-20} and
\astrobj{SGR 1900+14} showing giant flares are associated with
massive star clusters (Fuchs et al. 1999; Wachter et al. 2008),
and \astrobj{SGR 0525-66} is associated with a supernova remnant
(Gaensler et al. 2001; Eikenberry 2003). The best sites for
IceCube to search for high energy neutrinos from SGRs are active
star-formation regions in our galaxy, such as Westerlund 1, a
young massive star cluster, in which several neutron stars and
magnetars were discovered (Muno et al. 2006; Muno et al. 2007;
Clark et al. 2008). For electromagnetic signals, the detection
distance can be as far as $\sim 70 ~\rm Mpc$ by the 
Swift's Burst Alert Telescope (BAT), which gives an excellent opportunity
to observe extragalactic giant flares and hyperflares from SGRs
(Hurley 2005). Popov \& Stern (2006) proposed the most promising
targets for such observations, e.g., Virgo Cluster, NGC3256
etc. Although as a smaller detector, the Fermi's
Gamma-Ray Burst Monitor (GBM) is less sensitive than
BATSE and thus its detection distance for hyperflares is $\le$
40 Mpc, can augment the Swift's high energy
sensitivity. Via the Fermi's Large Area Telescope (LAT),
the detection can be naturally extended to the GeV
region (Gehrels et al. 2007). The broad-band studies will enable the
investigation of the hyperflare spectral and temporal
evolution over 7 orders of magnitude in wavelength, shedding light on the
mechanism of hyperflares. The gamma-rays decayed from the
neutral pions produced by hadronic interactions have a
flux and energy comparable to neutrinos. These TeV-PeV
gamma-rays might be detected by the southern Cherenkov
telescope, e.g., H.E.S.S, if the gamma-rays can escape the pair
production absorption (Kohnle et al. 2000).  

Besides possible detection of neutrinos from a single
hyperflare, the contribution to the neutrino background from
such hyperflares is also expected to be important since the
hyperflare explosion rate is high. The total energy spectrum
of the accelerated protons can be written as
\begin{eqnarray}
  \label{eq:diffuseflux}
  \epsilon_p^2\frac{dn_p}{d\epsilon_p}&\simeq&
  \frac{\xi_iL_{\rm kin}t_0}{{\rm
    ln}({\epsilon_{p,\rm max}}/{\epsilon_{p,\rm
      min}})}.
\end{eqnarray}
Then we can estimate the diffuse neutrino background flux from
hyperflares based on the template of the \astrobj{SGR 1806-20}
hyperflare by
\begin{eqnarray}
  e_{\nu}^2\phi_{\nu}&\sim&
  \frac{c}{4\pi H_0}\frac{1}{8}{\rm
    min}(1,f_{p\gamma})\epsilon_p^2\frac{dn_p}{d\epsilon_p}R_{\rm
    HF}(0)f_zf_b\nonumber\\
  &\simeq& {\rm min}(1,f_{p\gamma})\Big(\frac{R_{\rm
      HF}(0)}{90\times 365 {\rm~ GeV^{-3} ~yr^{-1}}}\Big)
  \Big(\frac{f_z}{3}\Big)(\frac{f_b}{0.1})\nonumber\\
  &&\times \cases{0.86\times 10^{-14}~\xi_{i,-0.5}~ L_{{\rm
        kin},45.5}~t_{0,-1} ~~({\rm Baryon Poor}) \cr
    0.86\times 10^{-11}~\xi_{i,-0.5}~ L_{{\rm kin},48.5}~t_{0,-1}~~({\rm Baryon Rich})}\nonumber\\
  &&{\rm GeV ~cm^{-2} ~s^{-1}~ str^{-1}}
\end{eqnarray}
where $H_0=71~{\rm km~s^{-1}~Mpc^{-1}}$, $f_b$ is the beaming
factor, $R_{\rm HF}(0)$ is the total hyperflare rate at $z=0$, and
$f_z$ is the correction factor for the possible contribution from
high redshift sources. Although we can always choose $\Delta t$ to
set that ${\rm min}(1,f_{p\gamma})=1$, the diffuse TeV-PeV
neutrino flux, however, also depends on the baryon loading of the
outflow, the explosion rate, the efficiency of energy conversion
and the geometry-corrected energy of the hyperflare, and
therefore it is more uncertain than the high energy neutrino flux
from a single source. Among these four parameters, the
efficiency $\xi_i$ is more stringently constrained because it is
less dependent on the model of hyperflares, the other three
parameters have large uncertainties due to low statistics with
using only one sample. In addition to strong hyperflares, SGRs
also show less intensive but more frequent giant flares, e.g., the
giant flares of March 5, 1979 from \astrobj{SGR 0525-66} with an
isotropic energy $E_{\rm iso}=3.6\times 10^{44}~\rm ergs$, and Aug
27, 1998 from \astrobj{SGR 1900+14} with $E_{\rm iso}=5.2\times
10^{43}~\rm ergs$. Their galactic rate is estimated to be about
$0.05-0.02~\rm yr^{-1}$ (Woods \& Thompson 2006), one order larger
than that of hyperflares. The less energetic giant flare may have
a larger beaming factor than typical hyperflares, thus the giant
flares and hyperflares would possibly provide the same
contributions to the diffuse neutrino background. Due to the low
statistics, it is unknown whether giant flares and hyperflares from
SGRs form a continuous luminosity distribution. We cannot
estimate the contribution of high energy neutrinos from
hyperflares with energy larger than the December 27, 2004 event.
Nevertheless, the baryon-rich model with high explosion rate and
large geometry-corrected energy predicts high diffuse neutrino
flux. To be specific, we estimate a TeV-PeV diffuse neutrino flux
of $4\times 10^{-11}~\rm GeV~cm^{-2}~s^{-1}~str^{-1}$ by using
an optimistic set of model parameters ($\eta=10$, $\xi_i=0.5$,
$f_b=0.25$, $L_{\gamma,47.5}=1.0$, $t_0=0.1$ and $R_{\rm HF}=100
~\rm Gpc^{-3}~ day^{-1}$). This value is about three orders of
magnitude higher than that estimated in magnetar steady phase (Zhang
et al. 2003), while more than one order of magnitude lower than that
from high-luminosity GRBs and low-luminosity GRBs (LL-GRBs) (Waxman \&
Bahcall 1997; Murase \& Nagataki 2006).

In addition to be served as high-energy neutrino factories,
hyperflares are also possible sources of CRs (Asano et al. 2006).
A baryon-rich outflow and a baryon-poor outflow can accelerate
protons to the maximal energies of $7\times 10^{18}$ eV and
$4\times 10^{16}$ eV, respectively (Ioka et al. 2005). Recently,
Wang et al. (2007) and Budnik et al. (2008) argued that the
trans-relativistic supernovae accompanied by LL-GRBs may be the
mecca of CRs within the energy range $10^{17}-10^{19}$ eV due to
their energetic outflows and high explosion rate. Is it possible
that hyperflare are important sources of CRs in this energy range?
For a baryon-poor outflow, low magnetic field can not accelerate
CRs to sufficient high energies and there is no adequate kinetic
energy available for CRs. While for a baryon-rich outflow, the
kinetic energy is $\sim 10^{47.5}$ ergs, four orders of magnitude
lower than that of LL-GRBs. On the other hand, the explosion rate
of hyperflares is two orders of magnitude higher than the latter.
So even if hyperflare have a baryon-rich origin, their
contribution to CRs within the energy range $10^{17}-10^{19}$ eV
is only $\sim1\%$ of that of LL-GRBs.

\section{Conclusions}
In this paper, we have calculated the diffuse TeV-PeV neutrino
flux from hyperflares of SGRs based on the template of the famous
2004 December 27 event and the fireball model. Because of a 
possible high explosion rate, albeit very uncertain, hyperflares
have a contribution, although likely less than $10\%$ of the
contribution from GRBs, to the diffuse high-energy neutrino
background. However, the nondetection of the neutrinos from
\astrobj{SGR 1806-20} by AMANDA II indicates that the hyperflare
may have a baryon-poor origin and thus its contribution to
neutrino background can be negligible. It is hard for IceCube to
detect the diffuse high energy neutrinos from hyperflares, though
which can put severe constraints on the hyperflare explosion rate.
However the neutrino detectors of IceCube and Km3NeT can
optimistically capture the TeV-PeV neutrinos from a single event
like the hyperflare by \astrobj{SGR 1806-20}, which may provide
independent clues on the trigger mechanism of the hyperflare in
addition to its electromagnetic emission. More observations are
required to improve the statistics and verify what scenario is
correct.

\section{Acknowledgments}
We would like to thank the anonymous referees' constructive
comments and suggestions which improved our paper significantly.
We thank Dr YWYu for improving the manuscript. This work was
supported by the National Natural Science Foundation of China
(grants 10473023, 10503012, 10621303, and 10633040), the National
Basic Research Program of China (973 Program 2009CB824800). XFW
also thanks the supports of NSF AST 0307376, NASA NNX07AJ62G,
NNX08AL40G, the Special Foundation for the Authors of National
Excellent Doctorial Dissertations of P. R. China by Chinese
Academy of Sciences, China Postdoctoral Science Foundation, and
Postdoctoral Research Award of Jiangsu Province.

\end{document}